\documentclass[a4paper]{jpconf}
\usepackage{graphicx}
\begin{document}
\title{Simulation of Quantum Universe}

\author{Sang Pyo Kim}

\address{Department of Physics, Kunsan National University, Kunsan 54150, Korea}

\ead{sangkim@kunsan.ac.kr}

\begin{abstract}
Quantum simulation provides quantum systems under study with analogous controllable quantum systems and has wide applications from condensed-matter physics to high energy physics and to cosmology. The quantum system of a homogeneous and isotropic field in the Friedmann-Robertson-Walker universe can be simulated by a charge in an electrically modulated ion trap. The quantum states of these time-dependent oscillators are constructed by quantum invariants. Further, we propose simulation of quantum Friedmann-Robertson-Walker universe with a minimal massive scalar field by a charged scalar field in a homogeneous, time-dependent, magnetic field in quantum electrodynamics and investigate the Cauchy problem of how the wave functions evolve.
\end{abstract}

\section{Introduction}

Understanding the evolution of quantum systems has been one of unsolved and challenging problems since the advent of quantum theory. In particular, the information about the evolution of a quantum system increases exponentially or geometrically with the degrees of freedom and requires huge data. Quantum simulation is an alternative to large or uncontrollable quantum systems and has wide applications from condensed-matter physics to high energy physics and to cosmology \cite{Georgescu:2013oza}. The advancement of technology in separating and confining a single charged or neutral particle within ion traps makes it possible precisely measure physical quantities with the unprecedented precision \cite{Paul:1990,Leibfried:2003zz}.

In gravity there have been many proposals to simulate or imitate nonperturbative nature of quantum fields in curved spacetimes.  An acoustic analogue of a black hole is the most well known model \cite{Unruh:1980cg}. A sonic black hole can be formed in Bose-Einstein condensates \cite{Garay:2000jj}
and the Unruh effect can be measured by an accelerating electron in intense lasers \cite{Schutzhold:2006gj}. Ion traps can be used to simulate and measure the Gibbons-Hawking radiation of a de Sitter space \cite{Menicucci:2010xs}.

Simulation of cosmology is another interesting arena because the homogeneous Bianchi models undergo Belinskii-Khalatnikov-Lifshitz's chaotic motions and infinite oscillations near the singularity (for a review and references, see refs. \cite{Belinski:1970,Belinski:1982} and also for numerical approaches, see  ref. \cite{Berger:1998us}). Though the Friedmann-Robertson-Walker (FRW) universe exhibits a simpler behavior near the singularity than the homogeneous Bianchi models, the quantum FRW cosmology with a minimal massive scalar field shows complicated and interesting motions that undergo infinite oscillations \cite{Kim:2012js,Kim:2013ixt} in contrast to the classical counterpart \cite{Belinski:1973}.

The main purpose of this paper is to propose simulations of quantum universe of the FRW geometry with a minimal massive scalar field both in the second quantized theory and the third quantized theory in the superspace of the geometry and scalar field. In the second quantized theory, the wave function of the universe obeys the Wheeler-DeWitt (WDW) equation, a relativistic wave equation, in the superspace of the three-geometry and the scalar field. In the third quantized theory, the WDW equation follows from an action of quantum fields in the superspace and the wave function has the status of operators that create or annihilate the universe of a given wave function just as a second quantized field creates or annihilates a particle and antiparticles. In the second quantized or third quantized theory, the universal interaction of gravity manifests itself through parametric interactions of the underlying geometry of general relativity. The quantum universe of the FRW geometry with massless fields is equivalent to a charged scalar field in time-dependent electric fields and that with a massive complex field to the charged field in time-dependent magnetic fields.

The organization of this paper is as follows. In section 2, we explore the quantum states of a charge modulated by time-dependent electric fields in an ion trap and compare those with the quantum system of a homogeneous and isotropic field (inflaton) in the FRW universe. By using linear quantum invariants acting as the annihilation and creation operators, the Fock space of all excited harmonic states can be constructed and the most general coherent-squeezed states and their dispersion relations can also be found. In section 3, we explain the third quantized theory of the FRW universe with a minimal massive complex scalar field and compare the theory with quantum electrodynamics (QED) of a charged scalar field in a homogeneous, time-dependent, magnetic field. The simulation of the quantized universe is proposed using a charged field in an electromagnetic field in QED. In section 4, we study the evolution of wave functions of the WDW equation and compare these with Landau levels in the time-dependent magnetic field.

\section{Ion Traps and Simulation of Quantum Fields in FRW Universe}\label{ion trap}

A Paul trap or an ion trap is a  device of generating two- or three-dimensional quadrupole electromagnetic fields that confine a charged or neutral particle \cite{Paul:1990,Leibfried:2003zz}. The charge with mass $m$ in properly arrayed and oriented fields has decoupled quadratic Hamiltonian
\begin{eqnarray}
H (t) = \sum_{i =1}^{3} H_{(i)} (t) := \sum_{i =1}^{3} \frac{1}{2m} p_{(i)}^2 + \frac{m}{2} \omega_{(i)}^2 (t) x_{(i)}^2, \label{osc ham}
\end{eqnarray}
where the frequencies squared are an interaction of the charge $q$ $(q > 0)$ with an electric potential
\begin{eqnarray}
\omega_{(i)}^2 (t) = \frac{q U_0}{m} \alpha_i + \frac{q f(t)}{m} \tilde{\alpha}_i.
\end{eqnarray}
Here, $x_{(i)}$ denotes $x, y$ and $z$, and $p_{(i)} = m \dot{x}_{(i)}$ and $f(t)$ is a modulating, time-dependent, electric field. The equations of motion are the harmonic motions with time-dependent frequencies squared
\begin{eqnarray}
\ddot{x}_{(i)} + \omega_{(i)}^2 (t) x_{(i)} = 0. \label{har mot}
\end{eqnarray}
The charge neutrality of the Paul trap demands the Gauss law, $\nabla^2 \Phi (\vec{x}, t) = 0$, which leads to the condition for any time
\begin{eqnarray}
\sum_{i =1}^{3} \alpha_i = 0, \qquad \sum_{i =1}^{3} \tilde{\alpha}_i = 0. \label{gauss con}
\end{eqnarray}
The condition (\ref{gauss con}) is the only constraint for, otherwise, decoupled oscillators (\ref{osc ham}). Depending on the parameters $\alpha_i$, $\tilde{\alpha}_i$ and the modulation field $f(t)$, at least one of frequencies squared can take a negative value.

The FRW universe with the metric
\begin{eqnarray}
ds^2 = - dt^2 + a^2 (t) d{\bf x}_3^2,
\end{eqnarray}
by redefining a new time variable $d \eta = dt/a^3 (t)$, can have another metric
\begin{eqnarray}
ds^2 = - a^6(\eta) d\eta^2 + a^2 (\eta) d{\bf x}_3^2. \label{new met}
\end{eqnarray}
For instance, a de Sitter space $a = e^{Ht}$, $\eta = - e^{-3 Ht}/3H$ has a range $(- \infty, 0)$.
Then the Hamiltonian for a complex scalar field with mass $\mu$ in the new metric is
\begin{eqnarray}
H_{\rm M} = \int \frac{d^3 {\bf k}}{(2 \pi)^3} \Bigl[ \pi^*_{\bf k} \pi_{\bf k} + \omega_{\bf k}^2 (\eta) \phi^*_{\bf k} \phi_{\bf k} \Bigr], \label{frw ham}
\end{eqnarray}
where
\begin{eqnarray}
\omega_{\bf k}^2 (\eta) = a^6 (\eta) \Bigl(\mu^2 + \frac{{\bf k}^2}{a^2 (\eta)} \Bigr).
\end{eqnarray}
The complex scalar field is equivalent to two real scalar fields. The equation of motion for each momentum is given by
\begin{eqnarray}
\ddot{\phi}_{\bf k} (\eta) + \omega_{\bf k}^2 (\eta) \phi_{\bf k} (\eta) = 0, \label{frw eq}
\end{eqnarray}
where dots denote derivatives with respect to $\eta$, and the same equation holds for $\phi_{\bf k}^*$. A homogeneous and isotropic field (inflaton) suppresses all momenta to ${\bf k} = 0$ and the Hamiltonian reduces to a quantum system of complex variable $\phi (\eta)$.

The exact quantum states for the time-dependent oscillators (\ref{osc ham}) and (\ref{frw ham}) can be found in terms of a quadratic quantum invariant by Lewis and Riesenfeld \cite{Lewis:1968tm}. Following refs. \cite{Kim:2001az,Kim:2003}, we use two linear quantum invariants that act as the annihilation and creation operators (units of $\hbar = c = l_{\rm P}^2 = 16 \pi/m_{\rm P}^2 = 1$ will be used)
\begin{eqnarray}
\hat{a}_{(i)} (t) &=& i \bigl[u_{(i)}^* (t) \hat{p}_{(i)} - m \dot{u}_{(i)}^* (t) \hat{x}_{(i)} \bigr], \nonumber\\
\hat{a}^{\dagger}_{(i)} (t) &=& - i \bigl[u_{(i)} (t) \hat{p}_{(i)} - m \dot{u}_{(i)} (t) \hat{x}_{(i)} \bigr], \label{an-cr op}
\end{eqnarray}
where $u_{(i)} (t)$ is a complex solution to eqs. (\ref{har mot}), (\ref{frw eq}) and (\ref{mathieu eq}) below that satisfies the Wronskian condition
\begin{eqnarray}
{\rm Wr} [u_{(i)} (t), u_{(i)}^* (t)] = \frac{i}{m}. \label{wron con}
\end{eqnarray}
Note that the complex solution is not unique since any linear superposition also satisfies eq. (\ref{wron con})
\begin{eqnarray}
u_{(i \nu)} (t) = \mu u_{(i)} (t) + \nu u_{(i)}^* (t), \qquad |\mu|^2 - |\nu|^2 = 1.
\end{eqnarray}
In quantum field theory, the choice of $u_{(i \nu)} (t)$ corresponds to the selection of the vacuum state: the vacuum constructed from $u_{(i \nu)} (t)$ is a squeezed vacuum of $u_{(i)} (t)$ and vice versa, as shown below.
The most general coherent(displaced)-squeezed states take the form \cite{Kim:2003}
\begin{eqnarray}
\Psi_{(i \nu) n} (x_{(i)}; x_{(i)c}, p_{(i)c}; u_{(i \nu)} (t)) &=& \Bigl(\frac{1}{\sqrt{2 \pi} 2^n n! |u_{(i \nu)}|} \Bigr)^{\frac{1}{2}} \Bigl(\frac{u_{(i \nu)}}{|u_{(i \nu)}|} \Bigr)^{n+ \frac{1}{2}} e^{i p_{(i)c} x_{(i)}} \nonumber\\&& \times H_{n} \Bigl(\frac{x_{(i)} - x_{(i)c}}{\sqrt{2} |u_{(i \nu)}| } \Bigr) \exp \Bigl[ i \frac{m}{2} \frac{\dot{u}_{(i \nu)}^*}{u_{(i \nu)}^*} \bigl( x_{(i)} - x_{(i)c} \bigr)^2 \Bigr], \label{cs har st}
\end{eqnarray}
where $H_n$ is the Hermite polynomial and the centroid of the wave packets are
\begin{eqnarray}
x_{(i)c} (t) &=& \langle \Psi_{(i \nu) n} (x_{(i)}; u_{(i \nu)} (t)) \vert \hat{x}_{(i)} \vert \Psi_{(i \nu) n} (x_{(i)}; u_{(i \nu)} (t)) \rangle, \nonumber\\
p_{(i)c} (t) &=& \langle \Psi_{(i \nu) n} (x_{(i)}; u_{(i \nu)} (t)) \vert \hat{p}_{(i)} \vert \Psi_{(i \nu) n} (x_{(i)}; u_{(i \nu)} (t)) \rangle.
\end{eqnarray}
Their dispersion relations are
 \begin{eqnarray}
\Delta x_{(i)}(t) &=& \langle \Psi_{(i \nu) n} (x_{(i)}; u_{(i \nu)} (t)) \vert \bigl( \hat{x}_{(i)} - x_{(i)c} (t)\bigr)^2 \vert \Psi_{(i \nu) n} (x_{(i)}; u_{(i \nu)} (t)) \rangle \nonumber\\ &=& (2n+1)   u^*_{(i \nu)} (t)  u_{(i \nu)} (t)\nonumber\\
\Delta p_{(i)} (t) &=& \langle \Psi_{(i \nu) n} (x_{(i)}; u_{(i \nu)} (t)) \vert \bigl( \hat{p}_{(i)} - p_{(i)c} (t) \bigr)^2 \vert \Psi_{(i \nu) n} (x_{(i)}; u_{(i \nu)} (t)) \nonumber\\ &=& (2n+1) m \dot{u}^*_{(i \nu)} (t)  \dot{u}_{(i \nu)} (t) \rangle. \label{dis rel}
\end{eqnarray}
The scalar field model (\ref{frw ham}) has $m = 1$ from eqs. (\ref{an-cr op}) to (\ref{dis rel}).

The Paul trap has a sinusoidal field $f(t) = \tilde{U}_0 \cos (\omega_{\rm rf} t)$ with a radio frequency to confine the charge.  In real experiments, a pure oscillating electric field with $\alpha_i = 0$ for $i =1, 2, 3$ and $\tilde{\alpha}_3  = -(\tilde{\alpha}_1 + \tilde{\alpha}_2)$ dynamically confines the charge while $\alpha_3 = -(\alpha_1 + \alpha_2) > 0 $ and $\tilde{\alpha}_1 = - \tilde{\alpha}_2$ provides also a dynamical confinement \cite{Leibfried:2003zz}. An experimental design of three-dimensional trap, $U_0 = 0 \sim 50\, {\rm V}$, $\tilde{U}_0 = 100 \sim 500\, {\rm V}$ and $\omega_{\rm rf}/ 2 \pi = 100\, {\rm kHz} \sim 100\, {\rm MHz}$ \cite{Leibfried:2003zz}, implies that at least one of the equations of motion has a negative frequency squared
\begin{eqnarray}
{x}''_{(i)} + \bigl[a_i - 2q_i \cos(2z) \bigr] x_{(i)} = 0, \label{mathieu eq}
\end{eqnarray}
where the prime denotes derivative with respect to $z$ and
\begin{eqnarray}
a_i = \frac{4 q U_0}{m \omega_{\rm rf}^2} \alpha_i , \qquad q = - \frac{2 q \tilde{U}_0}{m \omega_{\rm rf}^2}  \tilde{\alpha}_i, \qquad z = \frac{\omega_{\rm rf}}{2} t.
\end{eqnarray}
The Mathieu equation (\ref{mathieu eq}) has either stability regions or instability regions depending on parameters $a_i$ and $q_i$, whose stability regions lead to confinement of the charge \cite{Nist:2010}. Two independent real solutions with ${\rm Wr} [w_{(i)0}, w_{(i)1}] = 1$ combine to yield a complex solution to eq. (\ref{mathieu eq}) satisfying the Wronskian condition (\ref{wron con})
\begin{eqnarray}
u_{(i)} (t) = \frac{1}{\sqrt{2}} \bigl(w_{(i)0} (t) - i w_{(i)1} (t) \bigr).
\end{eqnarray}

An ion trap experiment was proposed to observe by the Unruh-DeWitt's monopole detector the analogue of quantum effects in cosmological spacetimes, such as the Gibbons-Hawking effect and inflationary structure \cite{Menicucci:2010xs}.
Phonon excitations of an ion in a trap with a trap frequency exponentially modulated with $\kappa$ exhibit a thermal spectrum with the Unruh temperature $T_{\rm U} = \kappa/ k_{\rm B}$ \cite{Alsing:2005dno}.

\section{Third Quantized Universe vs Second Quantized scalar QED} \label{3rd QU vs 2nd QED}

We study the quantum cosmology of the FRW geometry with a minimal complex scalar field $\phi$ with mass $\mu$, which has the extended supermetric
\begin{eqnarray}
ds^2 = - da^2 + a^2 d\phi^* d\phi = e^{2 \alpha} \bigl(- d \alpha^2 + d\phi^* d\phi \bigr). \label{sup met}
\end{eqnarray}
Then, the Hamiltonian constraint in the Arnowitt-Deser-Misner (ADM) formalism and the WDW equation take the form (for a review and references, see ref. \cite{Kiefer:2004gr})
\begin{eqnarray}
{\rm H} (\alpha, \phi, \phi^*) = - \underbrace{\bigl(\pi^2_{\alpha} + V_{\rm G} (\alpha) \bigr)}_{\rm gravity~part ~{\rm H}_{\rm G}} + \underbrace{\bigl(\pi_{\phi}^* \pi_{\phi} + \mu^2 e^{6 \alpha} \phi^* \phi \bigr)}_{\rm scalar~field~part~ {\rm H}_{\rm M}}
\end{eqnarray}
and
\begin{eqnarray}
\Bigl[ \opensquare + e^{6 \alpha} \mu^2 \phi^* \phi -  V_{\rm G} (\alpha) \Bigr] \Psi (\alpha, \phi, \phi^*) = 0, \label{WDW eq}
\end{eqnarray}
where the d'Alembertian in the superspace and the gravitational potential are, respectively,
\begin{eqnarray}
\opensquare =  \frac{\partial^2}{\partial \alpha^2} - \frac{\partial^2}{\partial \phi \partial \phi^*}, \qquad V_{\rm G} (\alpha)= k e^{4 \alpha} - 2 \Lambda e^{6 \alpha}. \label{wav op}
\end{eqnarray}
Here, $k = 1, 0, -1$ for a closed, flat and open universe and $\Lambda$ denotes a cosmological constant. In terms of the real and imaginary components $(\phi = \phi_1 + i \phi_2)$, the complex scalar field equals to two real scalar fields $\phi^*\phi = \phi_1^2 + \phi_2^2$ and the d'Alembertian becomes a three-dimensional wave operator
\begin{eqnarray}
\opensquare = \frac{\partial^2}{\partial \alpha^2} - \frac{\partial^2}{\partial \phi_1^2} - \frac{\partial^2}{\partial \phi_2^2}.
\end{eqnarray}
The WDW equation is a relativistic wave equation for scalar fields with $\alpha$-dependent mass $e^{3 \alpha} \mu$ and with an $\alpha$-dependent effective mass $V_{\rm G} (\alpha)$ for the field $\Psi$.

There have been a few approaches to understand and study the WDW equation (\ref{WDW eq}). One approach is to look for the wave function with specific boundary conditions necessary for describing the present universe. Hartle-Hawing's no-boundary wave function is such a wave function that matches the present Lorentzian geometry to a Euclidean compact manifold \cite{Hartle:1983ai,Hawking:1983hj} and thus satisfies a specific boundary condition.
The other approach is to interpret eq. (\ref{WDW eq}) as a relativistic wave equation for scalar fields with $\alpha$-dependent mass and an $\alpha$-dependent effective mass $V_{\rm G}$ in the superspace $(\alpha, \phi_1, \phi_2)$ with supermetric (\ref{sup met}). The Cauchy problem has been developed in refs. \cite{Kim:2013ixt,Kim:1992tc} that evolves the initial data at a given hypersurface $\Sigma = \alpha_0$ to any hypersurface $\Sigma = \alpha$ of the superspace in the same way as the ADM formalism does in classical general relativity.

Along the line of the second approach of relativistic wave equation, we may third-quantize the WDW equation in the superspace.
Then the WDW equation follows from the third quantized action
\begin{eqnarray}
S = \int d \alpha d\phi_1 d \phi_2 \Bigl[ - \Psi\frac{\partial^2}{\partial \alpha^2} \Psi + \Psi \Bigl(\frac{\partial^2}{\partial \phi_1} + \frac{\partial^2}{\partial \phi_2} \Bigr) \Psi - \Psi \Bigl( e^{6 \alpha} \mu^2 \bigl(\phi_1^2 + \phi_2^2 \bigr) - V_{\rm G} (\alpha) \Bigr) \Psi \Bigr]. \label{3rd act}
\end{eqnarray}
The super-Hamiltonian from the action (\ref{3rd act}) is
\begin{eqnarray}
H = \int d\phi_1 d \phi_2 \Bigl[ \Pi^2 + \Psi \Bigl(- \frac{\partial^2}{\partial \phi_1} - \frac{\partial^2}{\partial \phi_2} +  e^{6 \alpha} \mu^2 \bigl(\phi_1^2 + \phi_2^2 \bigr) \Bigr) \Psi - \Psi V_{\rm G} (\alpha) \Psi \Bigr], \label{sup ham}
\end{eqnarray}
where $\Pi = \partial \Psi/\partial \alpha$ is the conjugate to the wave function $\Psi$.
Massless fields lead to a super-Hamiltonian of $\alpha$-dependent oscillators \cite{Banks:1984cw,McGuigan:1988vi,McGuigan:1988es,Hosoya:1988aa,Abe:1993ye,Buonanno:1996um} and the tachyonic state was studied for the closed universe \cite{Kim:2012js}. We note that for the massless fields the wave function separates by $e^{i \pi_i \phi_i}$ $(i =1, 2)$ and the super-Hamiltonian (\ref{sup ham}) is an $\alpha$-dependent oscillator, which can be simulated by ion traps in section 2.

We now compare the WDW equation (\ref{WDW eq}) for the massive complex field and the Hamiltonian (\ref{sup ham}) with those of a charged scalar field in a magnetic field in scalar QED. We note that the third quantization of a massive field is analogous to the second quantized Klein-Gordon field with charge $q$ in a homogeneous, time-dependent, magnetic field
\begin{eqnarray}
{\bf A} (t) = \frac{1}{2} {\bf B} (t) \times {\bf x}. \label{sym gaug}
\end{eqnarray}
In fact, the transverse motion of a charged scalar field with mass $m$ in a temporal, homogeneous, magnetic field along the $z$-direction is given by \cite{Kim:2013lda}
\begin{eqnarray}
\Bigl[\frac{\partial^2}{\partial t^2} + \Biggl( {\bf p}^2_{\perp} + \Bigl(\frac{q B(t)}{2} \Bigr)^2 {\bf x}^2_{\perp} - q B (t) L_z\Biggr) + \bigl(m^2 + k_z^2 \bigr) \Bigr] \Phi_{\perp} (t, {\bf x}_{\perp}) = 0.
\end{eqnarray}
The Hamiltonian of the transverse part of the field has the form \cite{Kim:2014xxa}
\begin{eqnarray}
H_{\perp} (t) = \int d^2 {\bf x}^2_{\perp} \Bigl[\pi_{\perp}^* \pi_{\perp} + \Phi_{\perp}^* \bigl({\bf p}^2_{\perp} + \omega^2_L (t) {\bf x}^2_{\perp} - 2 \omega_L (t) L_z \bigr) \Phi_{\perp}  \Bigr]. \label{qed ham}
\end{eqnarray}
Thus the analog of quantum universe (\ref{sup ham}) and the second quantized scalar QED (\ref{qed ham}) is manifest:
\begin{eqnarray}
\nabla^2_{\perp} \Leftrightarrow \frac{\partial^2}{\partial \phi_1} + \frac{\partial^2}{\partial \phi_2}, \qquad \omega_L (t) \Leftrightarrow e^{3 \alpha} \mu, \qquad m^2 + k_z^2 \Leftrightarrow - V_{\rm G} (\alpha).
\end{eqnarray}
The difference due to $L_z$ becomes a technical point when the quantum state of the charged scalar and the wave function of the universe are expanded by the Landau levels and harmonic wave functions of the scalar fields, respectively, as will be shown in section 4.
Table 1 compares characteristics of quantum FRW universe with a minimal scalar and a scalar field in QED: massless scalar fields of quantum FRW universe correspond to a charged scalar field in a time-dependent electric field and a massive complex scalar to the charged scalar field in a time-dependent magnetic field.

\begin{center}
\begin{table}[h]
\caption{Quantum Simulation of Universe}
\centering
\begin{tabular}{@{}*{7}{l}}
\br
Quantum Universe & Scalar QED \\
\mr
Universes & Charged scalars \\
Wheeler-DeWitt equation & Klein-Gordon equation \\
Superspace of spacetime & Electromagnetic fields \\
Massive scalar in the early universe & Scalar in a time-dependent magnetic field \\
Coupling of harmonic wave functions & Coupling of Landau levels \\
Gravitational potential & Scalar in a time-dependent electric field \\
Wave functions of the universe & Quantum states of charged field \\
\br
\end{tabular}
\end{table}
\end{center}

\section{Landau Levels and Wave Functions of WDW Equation} \label{landau levels}

The Landau levels in a constant magnetic field are well known in nonrelativistic and relativistic theory. However, the time-dependency of the magnetic field makes Landau levels changing in time and instantaneous Landau levels are a good adiabatic approximation only for a slowly varying field. Another issue is the couplings of instantaneous Landau levels due to $L_z$, which can be diagonalized in some new Landau levels. The essential complication originates from the time-dependent Landau levels, which make the task of finding the exact Landau levels nontrivial in relativistic theory. The Cauchy problem has been proposed for the charged scalar field in a time-dependent magnetic field in the new Landau levels  \cite{Kim:2013lda}.

The WDW equation and the super-Hamiltonian contain the Hamiltonian of $\alpha$-dependent oscillators
\begin{eqnarray}
\hat{H}_{(i)} (\phi_i; \alpha) = - \frac{\partial^2}{\partial \phi_i^2} + e^{6 \alpha} \mu^2 \phi_i^2.
\end{eqnarray}
Each oscillator has the $\alpha$-dependent harmonic wave functions
\begin{eqnarray}
\hat{H}_{(i)} (\phi_i; \alpha) \vert \Phi_{(i)n} (\phi_i; \alpha) \rangle = e^{3 \alpha} \mu (2n + 1) \vert \Phi_{(i)n} (\phi_i; \alpha) \rangle.
\end{eqnarray}
However, the $\alpha$-dependency makes the eigenfunctions continuously transit to neighbor eigenfunctions preserving the parity \cite{Kim:2013ixt,Kim:1992tc}
\begin{eqnarray}
\frac{\partial}{\partial \alpha} \vert \Phi_{(i)n} (\phi_i; \alpha) \rangle = {\bf \Omega}_{(i)} \vert \Phi_{(i)n} (\phi_i; \alpha) \rangle,
\end{eqnarray}
where the transition matrix operator takes the form
\begin{eqnarray}
{\bf \Omega}_{(i)} := \frac{3}{4} \bigl(\hat{A}_{(i)}^2 - \hat{A}_{(i)}^{\dagger 2} \bigr).
\end{eqnarray}
Here, $\hat{A}_{(i)}$ ($\hat{A}_{(i)}^{\dagger}$) is an annihilation (creation) operator for the harmonic oscillator $H_{(i)}$ that lowers (raises) the quantum numbers by one unit.

The stratagem for the Cauchy initial value problem that evolves the wave function $\Psi (\alpha_0, \phi_1, \phi_2)$ to $\Psi (\alpha, \phi_1, \phi_2)$ is to expand the wave function by the eigenfunctions on each hypersurface $\Sigma = \alpha$
\begin{eqnarray}
\Psi (\alpha, \phi_1, \phi_2) = \sum_{l, n = 0}^{\infty} \psi_{l, n} (\alpha) \Phi_{(1)n} (\alpha) \rangle \Phi_{(2)n} (\alpha) \rangle,
\end{eqnarray}
and to solve the ordinary second-order differential equation for the multi-indexed column vector $\psi_{l, n} (\alpha)$ with multi-indexed off-diagonal transition matrices ${\bf \Omega}_{(i)}$ and the diagonal energy matrix ${\bf E} = 2 e^{3 \alpha} (\hat{A}_{(1)}^{\dagger} \hat{A}_{(1)} + \hat{A}_{(2)}^{\dagger} \hat{A}_{(2)} + 1 )$. The detailed procedure can be found in refs. \cite{Kim:2013ixt,Kim:1992tc}.

\section{Conclusion}

The quantum system of a charge in an electrically modulated ion trap is analogous to the quantum system of a homogeneous and isotropic scalar field (inflaton) in the FRW universe. Thus an ion trap that is properly modulated by a time-dependent electric field can simulate the quantum evolution of the inflaton in the universe. The WDW equation, the second quantized theory, for the FRW universe with a minimal complex scalar field and the super-Hamiltonian, the third quantized theory, in the superspace of the geometry and scalar field, are analogous to a charged scalar field in QED. Massless scalar fields correspond to the charged scalar field in a time-dependent electric field while a massive complex scalar field corresponds to the charged scalar field in a time-dependent magnetic field.

These analogies between a quantum field in the FRW universe and a charge in an electrically modulated ion trap, and also between the wave functions of the WDW equation and the quantum states of a charged scalar field either in time-dependent electric fields or magnetic fields open a new window for quantum simulation of the universe. Another challenging simulation of quantum universe is the homogeneous Bianchi models that exhibit chaotic motions and infinite oscillations \cite{Belinski:1970,Belinski:1982} and it would be interesting to see how a minimal scalar field changes the characteristic of the system.

\ack
The author would like to thank Christian Schubert for the warm hospitality at Instituto de F\'{\i}sica y Matem\'aticas (ifm),
Universidad Michoacana de San Nicol\'as de Hidalgo (UMSNH), where this paper was initiated, and Rong-Gen Cai for the warm hospitality at Institute of Theoretical Physics (ITP), Chinese Academy of Sciences (CAS), where this paper was completed. This work was supported by the Basic Science Research Program through the National Research Foundation of Korea (NRF) funded by the Ministry of Education (NRF-2015R1D1A1A01060626) and also in part by Institute of Basic Science (IBS) under IBS-R012-D1.

\end{document}